\documentclass[conference]{IEEEtran}

\usepackage{floatrow}
\usepackage{makeidx}  
\usepackage{amsmath}
\usepackage{amssymb}
\usepackage{graphicx}
\usepackage{comment}
\usepackage{color}
\usepackage{soul}
\usepackage{xspace}
\usepackage{caption}
\usepackage{subfig}
\usepackage{url}
\usepackage{float} 
\usepackage{tikz}
\usepackage{balance}

\newcommand*\dash{\unskip\kern.16667em---\penalty\exhyphenpenalty
	\hskip.16667em\relax}

{\makeatletter
	\gdef\xxxmark{%
		\expandafter\ifx\csname @mpargs\endcsname\relax 
		\expandafter\ifx\csname @captype\endcsname\relax 
		\marginpar{\textcolor{red}{xxx}}
		\else
		\textcolor{red}{xxx~}
		\fi
		\else
		\textcolor{red}{xxx~}
		\fi}
	\gdef\xxx{\@ifnextchar[\xxx@lab\xxx@nolab}
	\long\gdef\xxx@lab[#1]#2{{\bfseries [\xxxmark \textcolor{red}{#2}
			---{\scshape #1}]}}
	\long\gdef\xxx@nolab#1{{\bfseries [\xxxmark \textcolor{red}{#1}]}}
	\long\gdef\xxx@lab[#1]#2{}\long\gdef\xxx@nolab#1{}%

\newfloatcommand{capbtabbox}{table}[][\FBwidth]
\usepackage{blindtext}

\def\Sys{ScriptNet\xspace}
\def\MPL{LaMP\xspace}
\def\SCL{CPoLS\xspace}
\def\PIL{CPoLS\xspace}
\def\Emb{\textsc{Embedding}\xspace}
\def\LSTM{\textsc{LSTM}\xspace}
\def\MaxPool{\textsc{MaxPool1d}\xspace}

\def\RecConv{ \textsc{RecurrentConvolutions}\xspace}
\def\Conv{\textsc{Conv1D}\xspace}

\begin{document}
\title{\Large \bf \Sys: Neural Static Analysis for Malicious JavaScript Detection}

\author{\IEEEauthorblockN{Anonymous Submission to NDSS 2019}}

\author{\IEEEauthorblockN{Jack W. Stokes}
\IEEEauthorblockA{Microsoft Research\\}
\and
\IEEEauthorblockN{Rakshit Agrawal}
\IEEEauthorblockA{University of California, Santa Cruz\\}
\and
\IEEEauthorblockN{Geoff McDonald}
\IEEEauthorblockA{Microsoft Corporation}
\and
\IEEEauthorblockN{Matthew Hausknecht}
\IEEEauthorblockA{Microsoft Research}
}

\maketitle

\begin{abstract}
Malicious scripts are an important computer infection threat vector
in the wild. 
For web-scale processing, static analysis offers
substantial computing efficiencies. We propose the \Sys system for neural malicious JavaScript detection which is based on static analysis. We use the Convoluted Partitioning of Long Sequences (\PIL) model, which processes Javascript files as byte sequences. Lower layers capture the sequential nature of these byte sequences while higher layers classify the resulting embedding as malicious or benign. Unlike previously proposed solutions, our model variants are trained in an end-to-end fashion allowing discriminative training even for the sequential processing layers. Evaluating this model on a large corpus of 212,408 JavaScript files indicates that the best performing \PIL model offers a 97.20\% true positive rate (TPR) for the first 60K byte subsequence
at a false positive rate (FPR) of 0.50\%. The best performing CPoLS model significantly outperform several baseline models.
\end{abstract}

\section{Introduction}
\label{sec:intro}
%
%
The detection of malicious JavaScript (JS)
is important for protecting users against modern malware attacks.
Because of its richness and its ability to automatically run on most operating systems,
malicious JavaScript is widely abused by malware authors to infect users' computers and mobile devices.
JavaScript is an interpreted scripting language developed by Netscape that
is often included in webpages to provide additional dynamic functionality~\cite{JS}.
JavaScript is often included in malicious webpages, PDFs and email attachments.
To combat this growing threat, we propose \Sys, a novel deep learning-based system for the detection of
malicious JavaScript files.

\xxx[Jay]{With advances in browser and operating system
security making browser exploit attacks more difficult, adversaries are instead turning towards social engineering-based attacks.
In some cases, these attacks rely, either directly or indirectly, on the execution of malicious JavaScript.
Direct attacks include the execution of a JavaScript file which is attached in an email or the execution
of JavaScript code which is embedded in a webpage.  An indirect attack might involve the execution of JavaScript code which
is contained in a self-extracting archive file that is also found in an email attachment.}

There are numerous challenges posed by trying to detect malicious JavaScript.
Malicious scripts often include obfuscation to hide the malicious content
which unpacks or decrypts the underlying malicious script only upon execution.
Complicating
this is the fact that the obfuscators can be used by both benign and malware
files. Curtsinger \textit{et al.}~\cite{Zozzle} measured the distributions of malicious and
benign JavaScript files containing obfuscation. The authors showed that these distributions are very similar if a file is obfuscated and concluded that the presence of
obfuscation alone cannot be used to detect malicious JavaScript.
%

Another difficulty is that a large number of file encodings (\textit{e.g.}, UTF-8, UTF-16, ASCII)
are automatically supported by JavaScript interpreters. Thus, individual characters in the script may be encoded
by two or more bytes. As a result, malware script authors can use the embedding itself to attempt to hide malicious
JavaScript code~\cite{Xu2013}.

While a wide range of different systems
have been proposed for detecting
malicious executable files~\cite{Gandotra2014}, there has been less work in investigating malicious JavaScript.
Previous JavaScript solutions include those based purely on static analysis~\cite{Likarish2009,Shah2016}.
To overcome the limitations imposed by obfuscation,
other methods~\cite{Zozzle,Xu2013,Corona2014} include both static and some form of dynamic (\textit{i.e.}, runtime) analysis
to unroll multiple obfuscation layers.
In some cases, the solution is focused on the detection of JavaScript embedded in PDF documents~\cite{Laskov2011,Corona2014,Maiorca15}.
In addition, deep learning
models have recently been proposed for detecting system API calls in PE files~\cite{BenMalware,Kolosnjaji,PascanuMalware}, JavaScript~\cite{Wang2016},
and Powershell~\cite{Hendler2018}.

Including dynamic analysis allows improved detection over the
previous static analysis approaches.
While the combined static and dynamic analysis approaches can help unroll multiple obfuscation layers,
it can cause additional difficulties.

%
In some cases where latency or
a computational resources are problematic
we would like to have an effective, purely static analysis
approach for predicting if an unknown JavaScript file is malicious.
Three important applications of this work are large-scale, fast webpage, antimalware and email scanning services.
Search engine companies often scan large numbers of webpages searching for drive-by downloads.
Antimalware companies may scan hundreds of thousands or even millions of unknown files each day.
Similarly, large-scale email hosting providers often scan email attachments to identify
malicious content. To scan an individual webpage or file, a specially instrumented virtual machine (VM) must first be reset to a default configuration.
The webpage or email attachment is then executed, and dynamic analysis is used to determine whether
the unknown script makes any changes to the VM. This process is time consuming
and can require vast amounts of computing resources for extremely large-scale email services. If
a script classifier can be trained to accurately predict that a script attachment is benign based solely
on fast static analysis, this could possibly allow search and email service providers to reduce the number of expensive dynamic analysis performed
using full instrumented VMs in the cloud.
In this study, we focus on identifying malicious JavaScript for the Microsoft Windows Defender antimalware service.

To address these challenges, \Sys employs a sequential, deep learning model for the detection of malicious JavaScript files
based solely on static analysis. The deep learning system allows high accuracy even in the presence of obfuscation.
We evaluate the CPoLS sequence learning model family in the context of fully static analysis which is capable of capturing malicious behavior in any kind of JavaScript file. 

Since the system operates directly on the byte representation of characters instead of keywords, it is able
to handle the extremely large vocabulary of the entire script instead of detecting only the key API calls~\cite{Laskov2011,Maiorca15}.
In \Sys, a Data Preprocessing module first
translates the raw JavaScript files into a vector sequence representation.
The Neural Sequential Learning module then applies deep learning methods on the vector sequence
to derive a single vector representation of the entire file.
In this module, we propose a novel deep learning model model called
Convoluted Partitioning of Long Sequences (\SCL).
These models can operate on extremely long sequences and can learn a single vector representation of the input.
The next module of \Sys, the Sequence Classification Framework, then performs binary classification
on the derived vector and generates a probability $p_{m}$ of the input file being malicious.
Unlike earlier sequential models that are proposed to detect malicious PE files~\cite{BenMalware},
our models are trained
with end-to-end learning where all the model parameters 
are learned simultaneously
taking the JavaScript file directly as the input.

Evaluating the proposed models on a large corpus of 262,200 JavaScript files,
we demonstrate that the best performing \PIL
model offers a true positive rate of 97.20\% for the first 60K byte subsequences 
at a false positive rate of 0.50\%. 
%
We summarize the primary contributions of this paper as follows:
\begin{itemize}

	\item
	A comprehensive definition of a modular system is provided for detecting
	the malicious nature of JavaScript files using only the raw file content.
	
	\item
	A novel deep learning model is proposed for learning from extremely
	long sequences. 
	
	\item
	Strong malware detection results are demonstrated using \Sys on a large corpus of JavaScript
	files collected by hundreds of millions of computers running a production antimalware product.
	The results show the robustness of \Sys on predicting the malicious nature of JavaScript files
	that were obtained in the future and were not known at the time of training.
	
\end{itemize}




\section{Data Collection and Dataset Generation}
\label{sec:data}
Large labeled datasets are required to sufficiently train deep learning systems, and constructing a dataset
of malicious and benign scripts for training \Sys's models is a challenge.
When unknown JavaScript is encountered by the user during normal activity, it is submitted to the antimalware engine for
scanning.
Our antivirus
partner generated the datasets utilized in this study from JavaScript encountered by the Windows Defender antimalware engine
which was submitted to their production file collection and processing
pipeline.

\noindent
\textbf{Methodology:}
Entire JavaScript files are extracted from the incoming flow of files input to the production pipeline.
The antimalware engine is the only source of these files in this study which are uploaded from hundreds of millions of end user computers. A
user must provide consent (\textit{i.e.}, opt-in) before their file is transmitted to the production
cloud environment.
In many cases,
JavaScript files may be extracted
from installer packages or archives which are also processed by the antimalware engine and input to the production pipeline.
Because we want to model how well \Sys would perform in a production setting, we do not artificially insert additional
file into the datasets such as those collected from Alexa top 500 websites or by other data augmentation methods because the
production pipeline does not currently spend computational resources doing so.

\noindent
\textbf{Labels:} Similar to the raw script content, the labels are also provided by the antimalware company,
and the labeling process which is used in production cannot be changed for our study. We now describe
the labeling process that was used by the antimalware company
to generate the labels.

A script is labeled as malware if it has been inspected by our AV partner's analysts and determined to be malicious. In addition, the script is labeled as malicious if it has been detected by the company's detection signatures.
Finally, a JavaScript file is
labeled as malware if eight or more other anti-virus vendors detect it as malware. Given the huge volumes of files
that are scanned each day, it is possible that an individual anti-malware company may mispredict in an unknown file
is malicious or benign. The threshold parameter of eight was empirically chosen by the company after carefully selecting it
as a good tradeoff between identifying malicious scripts while minimizing false positives.

A script
is labeled as benign by a number of methods. First, the script is
considered benign if it has been labeled as benign by an analyst or
has been previously collected by a trusted source such as being downloaded
from a legitimate webpage or signed by a trusted signer. However, if this does not provide enough
labeled benign scripts, the dataset is augmented with lower confidence benign
scripts in the production pipeline which are not detected by the company's scanners and cloud detections
as well as by any other trusted anti-virus vendors for at least 30
days after our AV partner has first encountered it in the wild.

\noindent
\textbf{Datasets:}
As described in Table~\ref{tab:dataset_stats}, our anti-virus partner provided the full content of  262,200 JavaScript
files which contained 222,235 malicious and 39,965 benign scripts. For this research,
JavaScript files were subsampled from the production pipeline from September 2017 through
March 2018. These JavaScript files were partitioned into training, validation, and test sets
containing 151,840, 45,251, and 65,109 samples, respectively, based
on the non-overlapping time periods denoted in the table. 

For the learning phase, which is described later in the paper, we use the training and validation sets.
The training set is the largest portion ($\sim 60\%$)
and is used to train the learning model and update its weights.
The validation set is a small portion ($\sim 15\%$),
and is used for model parameter tuning during the learning phase.
JavaScript files in the validation set are not present in the training set.
The performance of the learning model on the validation set helps guide the selection of the best model.
In the detection phase, we use the third partition, called test set ($\sim 25\%$). 
JavaScript files in the test set are not present in either training or validation set.
The test set, consisting of new files, helps perform true evaluation of a trained model
on unseen data.
All the evaluation metrics for our models use the test dataset only.

\begin{table}[tb]
\begin{center}
\begin{tiny}
\begin{tabular}{| c | c | c | c | c | c | c | c |}
\hline
DataSet & Start & End & Total & Num  & \%   & Num  & \%  \\
        & Date & Date & &  Malware & Malware &  Benign & Benign \\
\hline
\hline
Training & 09/14/2017 & 12/28/2017 & 151,840 & 126,505 & 83.31 & 25,335 & 16.69 \\
Files & &  &  &  &  &  &  \\
Validation & 12/29/2017 & 02/01/2018  & 45,251 & 38,693 & 85.51 & 6,558 & 14.49 \\
Files & &  &  &  &  &  &  \\
Test & 02/02/2018  & 03/03/2018 & 65,109 & 57,037 & 87.60 & 8,072 & 12.40 \\ 
Files & &  &  &  &  &  &  \\  \hline
Total & 09/14/2017  & 03/03/2018 & 262,200 & 222,235 & 84.76 & 39,965 & 15.24 \\
Files & &  &  &  &  &  &  \\
\hline
\end{tabular}
\end{tiny}
\end{center}
\caption {DataSet Statistics.}
\label{tab:dataset_stats}
\end{table}

\section{\Sys System Design}
\label{sec:system}

\Sys is motivated by the objective of building a system, which can predict the malicious nature of a script by analyzing the file
in the absence of any additional information.
Additionally, we want this system to be able to learn features from the data itself without intervention from humans.
In this section, we describe the architecture of \Sys in detail which is designed to achieve these objectives.

\begin{figure*}[t]
	\subfloat[\Sys Training System Architecture]{
		\includegraphics[width=1\linewidth]{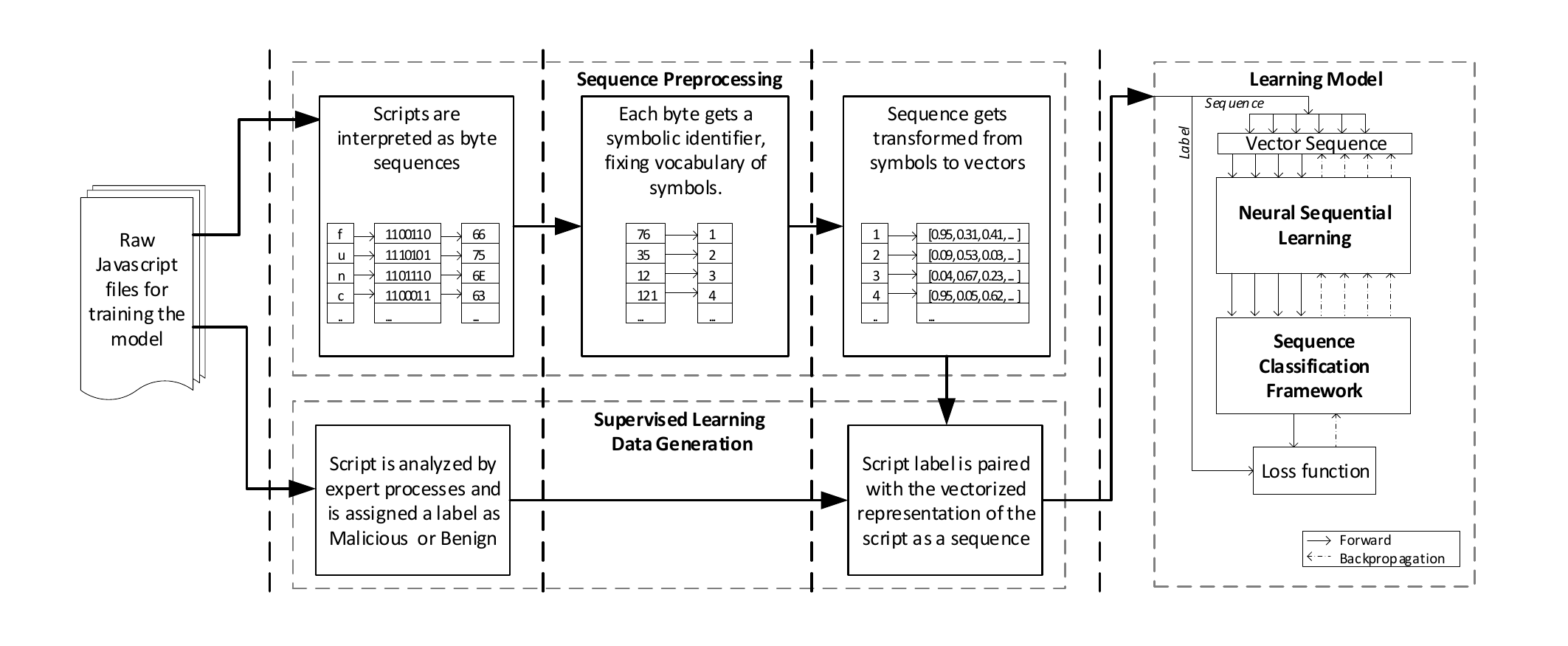}\label{subfig:system_training}
	}\\[-2em]
	\subfloat[\Sys Inference Pipeline]{
	\includegraphics[width=1\linewidth]{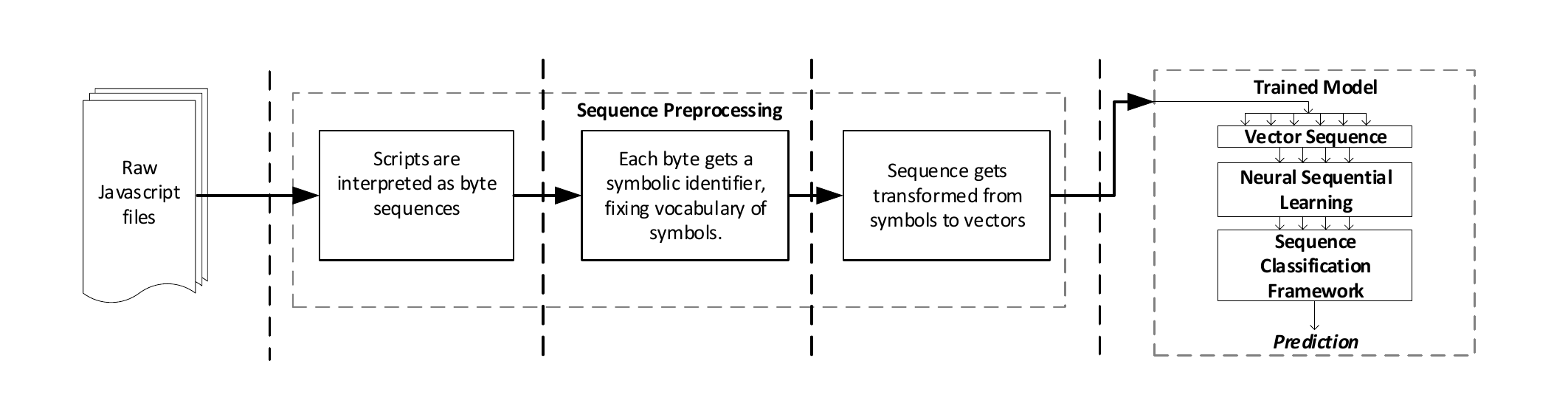}\label{subfig:system_inference}
	}
	
	\caption{\Sys system architecture for training and inference phases.}
	
	\label{fig:system}
\end{figure*}

The \Sys system is comprised of multiple modules banded together in a specific order.
These modules can be treated independently as black boxes that take a specific kind of input and can generate
a certain output, which can be used by the following module.
The high-level illustration of the \Sys system is shown in Figure~\ref{fig:system}.
In this section we describe the modules, as well as the process of learning and detection used by this system.

%
%
%
%

\noindent\textbf{Data Preprocessing:}
The first stage of \Sys is to process the raw file data and prepare it for utilization by a deep learning model.
In their raw form, the script files are simply text files written using readable characters.
The content inside script files is in the form of programming code.
This means that the text includes operators, variable names, and other syntactical properties.
In natural language, the semantic meaning of a certain word is limited to a small space.
Whereas in programming code, words cannot be directly mapped to a limited semantic space.
Moreover, the number of words and operators can grow infinitely as variable names do not need to follow any
linguistic limitations.
Therefore, we need to represent the scripts at a much finer level than using words.

\begin{figure}[tbh]
	\centering
	{\label{}\includegraphics[width=1\columnwidth]{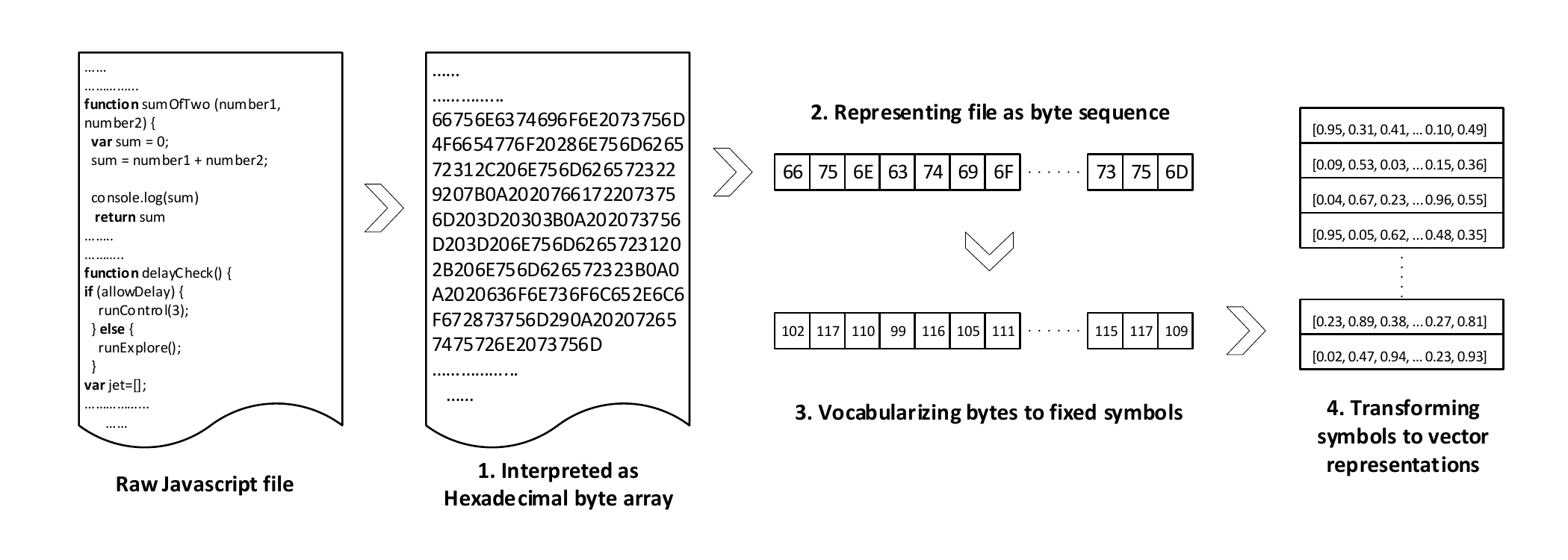}}
	\caption{Overview of the Sequence Processing Module of \Sys}
	\label{fig:sequence_processing}
\end{figure}

We achieve this by interpreting the script files as byte sequences.
Using this method to read the files,
we limit the space of possible options to the number of different bytes, i.e., $(2^8) = 256$.
The complete process of sequence processing is also illustrated in Figure~\ref{fig:sequence_processing}.

For the system to clearly identify the different bytes, we need to provide them unique identifiers.
At this stage, therefore, we create an index to map each byte with a symbolic identifier.
Following the convention in deep learning models,
we refer to this index as the vocabulary $V$ for the model.
A vocabulary helps maintain a one-to-one symbol mapping with the raw data
and can also be used to tie any out-of-vocabulary items to a special symbol. For example,
we introduce an extra symbol for padding sequences to a uniform length.

With the use of this vocabulary, we can now transform the input file into a sequence usable by the learning model.
At first, by reading the text as bytes, we get a byte sequence.
Next, we perform a lookup through the vocabulary index and represent each byte with its symbolic representation.
We refer to the derived sequence as $B$, where $B = [b_1, b_2, b_3, \ldots] \hspace{0.5em} \forall b_i \in V$
denotes a sequence of symbols $b_i$ each of which is identified in our vocabulary $V$.

For learning purposes, it is possible to directly use this symbolic representation.
However, symbols serve information at a very low level of dimensionality.
When represented as symbols, any similarity between two kinds of bytes cannot be directly identified.
In neural networks, the concept of representations, or 'embeddings' is extensively used for this purpose.
By representing symbols with vectors, we can increase the dimensionality of the information associated
with each element.
These vectors can be learned from the data itself.
The distance between these vectors also serves as a measure of semantic similarity.

In our case, we use this concept of representations and transform the symbolic sequence
into a sequence of vectors.
The initial value of these vectors is randomly selected using initialization methods by Glorot and Bengio~\cite{glorot10a}.
During the training phase, the vectors are updated along with the model.

\noindent\textbf{Neural Sequential Learning:}
In our preprocessing phase, we converted the input files into vector sequences.
Since the lengths of these files can vary, the derived sequences are also of different length.
General learning methods based on feature vectors read fixed length vectors as input and operate on them.
For our case with more complex-shaped data, we need to use modules that can process two-dimensional
input data.
Therefore, to learn from sequences, we use advanced neural network architectures like
Recurrent Neural Networks (RNNs) and Convolutional Neural Networks (CNNs)~\cite{LeCun1995ConvolutionalNF}.

RNNs operate on each vector sequentially
and use both the input at each time step of the sequence,
along with the learned vector from the previous time step,
to produce the next output.
In our models, we use a memory-based variant of RNNs known as Long Short-Term Memory (LSTM)~\cite{Hochreiter1997,GersJj1999} neural networks.

CNNs operate by performing convolutions over a sliding window of smaller partitions within the
input.
These can operate on both sequences, as well as images.
For learning from sequences, we use one-dimensional CNNs.
While mostly used in computer vision~\cite{krizhevsky2012imagenet,russakovsky2015imagenet},
CNNs have also recently shown success in sequential learning~\cite{Gehring2016,Gehring2017}.

In our system, we construct Neural Sequential Learning modules based on both LSTMs and CNNs.
The objective of this module is to capture variable-length vector sequences
and learn a vector representation from them.
Intuitively, this module is responsible for searching through the input sequence
and extracting any relevant information that can be used for final detection.
In this paper, we use the Convoluted Partitioning of Long Sequences (\SCL) for sequence learning.
We will describe this model in detail in the next section.


\noindent\textbf{Sequence Classification Framework:}
Once the input file has been processed through our data processing and neural sequence learning modules,
the input is available as a fixed-length vector $h_{CL}$.
The objective at this stage is to perform the final prediction in order to classify the input as being
\emph{malicious} or \emph{benign}.
There are several methods of classification in machine learning that can be used at this point.
%
Since we use a sequence learning module based on neural networks,
we use classification models that can also be trained
using gradient descent-based methods.
Using such models, we can train our entire learning system end-to-end.
End-to-end learning means that every weight (or coefficient) in our model can be trained in a single process
guided directly by the ground truth.

\noindent\textbf{Learning Phase:}
The modules described above, combined together, create the complete \Sys system.
Since these models depend on data for training,
the system goes through a learning phase before it can be used on new files for malware detection.
Figure~\ref{subfig:system_training} describes the learning phase of \Sys.
In this phase, along with the JavaScript files, we also have labels associated with each file,
specifying them as being either \emph{malicious} or \emph{benign}.
The system, therefore, processes the input file into a vector sequence, as well as generates the associated label,
to be used for training the system.
The sequence vector-label pair is then passed to the learning model.

As mentioned above, we use gradient descent-based methods to train our models.
In such methods, a loss $\mathcal{L}$ is measured by comparing the prediction $p_m$ generated by the learning model
and the available ground truth label $\tau$.
This loss is then used to update the coefficients (\textit{i.e.}, weights) of the model.
For our objective of binary classification, we use the \emph{binary cross-entropy} loss function, which is defined as:
\begin{equation}
	\mathcal{L} = -(\tau log(p_m) + (1-\tau)log(1-p_m))
\end{equation}
where $\tau$ is the known ground truth, $p_m$ is the predicted probability of maliciousness, and
$\log$ is the natural logarithmic function.
%
This process of backpropagating the loss is repeated for the entire training dataset, for a specified number of iterations,
until the model converges to the best weights.


\noindent\textbf{Detection Phase:}
Once the model is trained completely,
it is then available for performing detection on new, unknown files.
In the learning phase, we use each sample to improve the learning model.
Whereas in the detection phase, we use a well-trained model to perform inference in a real deployment setup.
Figure~\ref{subfig:system_inference} shows the detection phase version of \Sys.

Since the learning model takes a vector sequence as an input,
the same data preprocessing module is used in the detection phase.
However, the model is now used only to generate a probability of detection.
In machine learning terminology, this pass through the model
is known as a forward pass.
During a forward pass, we only move through the model in one
direction and generate a probability $p_m$ for the input file.


\noindent\textbf{End-to-End Learning:}
Due to the modular nature of our system, we have the freedom to train it in different ways.
While we present neural models in this paper, the system can also use different
components from machine learning.
For instance, in the Sequence Classification Framework, we can ideally use any classifier
like an Support Vector Machine or Naive Bayes,
which may or may not support gradient-based updates.

By keeping our models in the realm of neural networks,
we are also able to utilize the concept of end-to-end learning.
This means that our system can train itself completely by just using the input files
and labels.
We do not need to train different modules individually in such a setting.
The results presented in this paper were trained using end-to-end models.

\section{Models}
\label{sec:model}

\Sys uses end-to-end learning models based on neural networks.
For our objectives, we need models for sequential learning
and sequence classification.
In this section, we present a detailed description of the
Convoluted Partitioning of Long Sequences (\SCL) model.
We first briefly discuss the neural method of sequence learning
and describe our motivation behind constructing these models.

\subsection{Sequence Learning and Limitations}
Learning from sequential data is a common use case in machine learning.
Data in natural language, speech, time series, stock prediction, and many
other domains can be of sequential nature.
For sequences of very short fixed lengths,
vector based learning models like logistic regression can often work well,
by flattening the sequence into a longer single vector.
For longer sequences, RNNs, specifically LSTMs have been popularly used
and have shown exceptional results.
Since CNNs can also capture multi-dimensional data,
they are often used with sequential data.

As the length of the sequences keeps increasing,
these models start experiencing different challenges.
RNNs originally experienced the vanishing and exploding gradient~\cite{BengioLongTerm1994,Hochreiter1998} problems
when used on longer sequences.
LSTMs, in particular help mitigate such problems.
However, for extremely long sequences,
models directly based on the LSTM can quickly become computationally expensive
and are unable to process the complete input.
Additionally, the objective of \Sys is to detect the malicious nature within a file.
In natural language data, sequences often generate a context space,
within which the semantic relationship between different objects is more clear.
For instance, in a task of detecting abusive text, a sentence containing foul words
is also semantically linked to an abusive sentiment.
The mere presence of a foul word in a sentence cannot make it abusive.
However, in malicious JavaScript files,
a file can look completely normal except for a certain point in the file where malicious code
is present.
Therefore, not only do we need to process very long files,
we also need to capture specific malicious nature hidden at any location within the file.

Due to the limitations mentioned above and problem specific requirements, we propose these new models.
The \SCL model
can operate on extremely long sequences.
We later show in the paper, that these models also perform exceedingly well over other proposed
solutions in a similar problem space.

\subsection{Convoluted Partitioning of Long Sequences}
Convoluted Partitioning of Long Sequences (\SCL) is a neural model architecture designed specifically
to extract classification information hidden deep within long sequences.
In this model, we process the input sequence by splitting it into smaller parts of fixed-length,
processing them individually, and then combining them again for further learning.
The process of \SCL is as follows:

\textbf{Step 1.}:
The model receives an input sequence $B$ as a sequence of bytes.
Since the sequences are extremely long, we pass them to the model in their symbolic form
and transform them to vectors later.

\textbf{Step 2}:
The byte sequence $B$ is split into a list $C$ of small subsequences $c_i \in C$ where
$i$ is the index of each partition in $C$.
During the split, the subsequences maintain their order.

\textbf{Step 3}:
Next on the smaller subsequences, we perform the lookup for transforming symbolic sequence $c_i$
into vector sequences $e_i \in E$ where $E$ is the list of vector subsequences
and $i$ is the index of each subsequence.
Following the conventional neural network terminology,
we refer to the layer for this lookup as the \Emb layer.

\textbf{Step 4}:
Each of these partitions $e_i$ are now separately processed through a module called \RecConv,
while still maintaining their overall sequential order.

\textbf{Step 4.1}:
In \RecConv, we pass each partition $e_i$ through a one-dimensional CNN, \Conv,
which applies multiple filters on the input sequence and generates a tensor $e^\chi_i$
representing the convoluted output of vector sequence $e_i$.
The combined list of convolved partitions $e^\chi_i$ for each subsequence $e_i \in E$
is referred to as $E^\chi$.

\textbf{Step 4.2}:
We then reduce the dimensionality of $e^\chi_i$  by performing a
temporal max pooling operation \MaxPool on it.
\MaxPool takes a tensor input $e^\chi_i$ and extracts a vector $e'_i$ from it
corresponding to the maximum values across each dimension.

\textbf{Step 5.}
As a result of \RecConv, for each subsequence $e_i$, we derive a vector
representation $e'_i$.
We finally combine these vectors in order to generate a new vector sequence
$E'$ where each vector $e'_i \in E'$ is a result of \RecConv
and, therefore, consists of the learned information from subsequence $e_i$.

\textbf{Step 6.}:
We now obtain a reduced-length sequence of vectors $E'$.
This sequence can now be processed using a standard sequence learning approach.
We, therefore, next pass this sequence through an LSTM.
In place of LSTM, we can also use multiple stacked LSTMs, bi-directional LSTMs (BiLSTMs),
or any other RNN variants like Gated Recurrent Units (GRUs), etc.
For an input sequence $E'$ of length $n$,
this layer produces a learned sequence $H_L$ of length $n$ but with a different fixed dimensionality.

\textbf{Step 7.}:
For detecting malware, we want to obtain the important malicious signal information within
the sequence $H_L$.
An effective method for such cases is the use of temporal max pooling, \MaxPool,
as proposed by Pascanu \textit{et al.}~\cite{PascanuMalware}.

Given an input vector sequence  $S = [s_0, s_1,\ldots s_{M-1}] \in S$ of length $M$,
where each vector $s_i \in \mathbb{R}^K$ is a $K$-dimensional vector,
\MaxPool computes an output vector $s_{MP} \in  \mathbb{R}^K$ as
$s_{MP}(k) = \max(s_0(k), s_1(k), \cdots s_{M-1}(k)) \forall k \in K$.
The vector $s_{MP}$, therefore, for each dimension, contains
the maximum value observed in the sequence for that dimension.

At this stage,
we pass the sequence $H_L$ through \MaxPool to obtain the final vector $h_{CL}$.
The vector $h_{CL}$ is the derived vector representation of the entire
sequence $B$ using the \SCL model.
This vector can now be used by the Sequence Classification Framework to perform
the final binary classification.

The simplest such model can be a logistic regression model that uses $h_{CL}$ and derives
a probability of maliciousness $p_m$.
We can even use complex models,
such as feed-forward neural networks, or deeper neural networks with multiple layers for the same purpose.
We can also use any non-linear activation functions in these networks.
For our experiments, we use the Rectified Linear Unit (ReLU),
which is defined as:
\begin{equation}
	f(x) = max(0,x)
\end{equation}
where $f$ represents the ReLU function on an input $x$.

Due to the modular nature of our system, the choice of the classifier is independent of the
sequential learning method being used.
For this reason, we evaluated our models with a large number of combinations.
Along with \SCL and \PIL for sequential learning,
we also used a simpler LSTM-based method.
We refer to this method as using LSTM and Max-Pooling (\MPL) based on the malware detection model proposed by Athiwaratkun and Stokes~\cite{BenMalware}.
As the name of the model suggests, this model  directly takes the vector input sequence,
and passes it through the  LSTM.
The sequence of learned vectors from the LSTM is then passed through a temporal max-pooling layer,
\MaxPool, in order to derive the final vector $h_{CL}$.
For an input byte sequence $B$, \MPL can be summarized as:
\begin{equation}
\begin{split}
& E = \Emb(B)\\
& H_L = \LSTM(E) \\
& h_{CL} = \MaxPool(H_L)
\end{split}
\end{equation}
where \Emb is the embedding lookup layer,
$E$ is the vector sequence derived using \Emb,
and $H_L$ is the output vector sequence derived from the LSTM.

\begin{figure*}[tbh]
	\centering
	{\label{}\includegraphics[width=0.8\columnwidth]{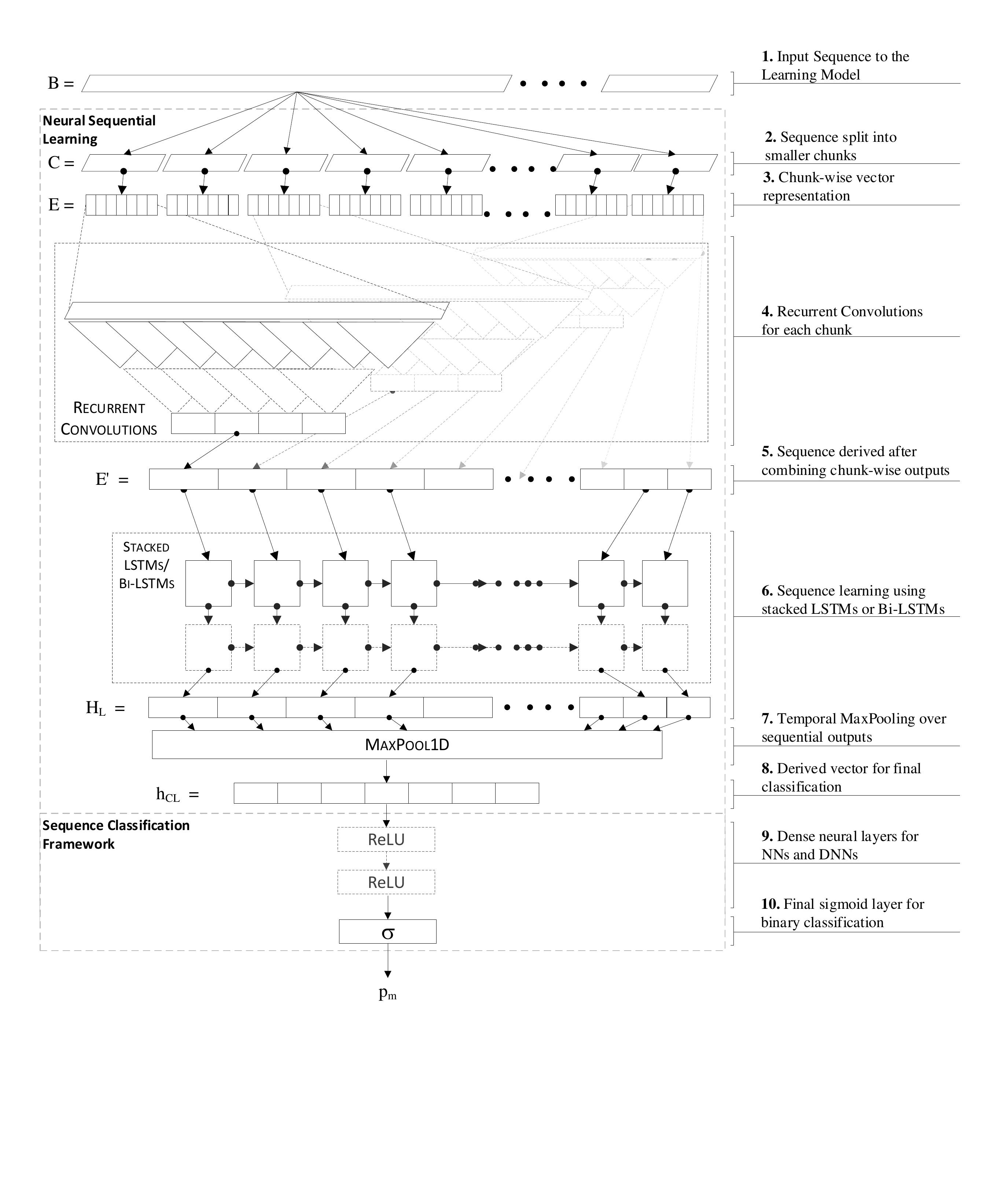}}
	\caption{Model for Convoluted Partitioning of Long Sequences (\SCL).}
	\label{fig:scl}
\end{figure*}

\section{Experimental Results}
\label{sec:eval}
In this section, we evaluate the performance of the proposed \Sys classifier models on the JavaScript
files described in Section~\ref{sec:data}.
We start by describing the experimental setup used to generate the results. We next investigate the performance of the different
hyperparamer settings for the \PIL
variants.
\noindent\textbf{Experimental Setup:}
All the experiments are written in the Python programming language using the Keras~\cite{keras} deep learning library with TensorFlow~\cite{Tensorflow} as the backend deep learning
framework.
The models are trained and evaluated on a cluster of NVIDIA P100 graphical processing unit (GPU) cards.
The input vocabulary
size is set to 257 since the sequential input consumed by each model is a byte stream, and an additional symbol is used for padding shorter sequences within each minibatch.
All models are trained using a maximum of 15 epochs, but early stopping is employed if the model
fully converges before reaching the maximum number of epochs. The Adam optimizer~\cite{Kingma2014} is used to train all models.

We did hyperparameter
tuning of the various input parameters for
the JavaScript models, and the final settings are summarized in
Table~\ref{tab:hyper}. To do so, we first set the other
hyperparameters to fixed values and then vary the hyperparameter under consideration. The best parameter setting
is then set based on the validation error rate.
For example, to evaluate different minibatch sizes for the JavaScript LaMP classifier, we first set
the LSTM's hidden layer size $H_{LaMP} = 1500$, the embedding dimension to $E_{LaMP} = 50$, the number of LSTM layers $L_{LaMP} = 1$ and
the number of hidden layers in the classifier $C_{LaMP} = 1$. With these settings, we evaluate the classification error rate
on the validation set for the JavaScript dataset.

The \PIL model
is designed to operate on the full JavaScript sequences. However, training on the full
length sequences exhausts the memory capacity of the NVIDIA P100s in our cluster, depending on the particular variant and parameter settings of the model.
To overcome this limitation, we truncated the sequence length to $T$ = 60,000 bytes for all the \PIL and \SCL experiments.
Similarly, we truncated the sequences to
lengths of $T=200,1000$
bytes for the LaMP and Kolosnjaji CNN~\cite{Kolosnjaji} baselines.

\begin{table}[tb]
\begin{center}
\begin{scriptsize}
\begin{tabular}{| c | c | c | c |}
\hline
Model & Parameter & Description & Value \\
\hline
\hline
CPoLS & $T_{CPoLS}$  & Maximum Sequence Length & 60,000 \\
CPoLS & $B_{CPoLS}$  & Minibatch Size & 50 \\
CPoLS & $H_{CPoLS}$  & LSTM Hidden Layer Size & 250 \\
CPoLS & $E_{CPoLS}$  & Embedding Layer Size & 100 \\
CPoLS & $W_{CPoLS}$  & CNN Window Size & 10 \\
CPoLS & $S_{CPoLS}$  & CNN Window Stride & 5 \\
CPoLS & $F_{CPoLS}$  & Number of CNN Filters & 100 \\
CPoLS & $D_{CPoLS}$  & Dropout Ratio & 0.5 \\
\hline
LaMP & $T_{LaMP}$  & Maximum Sequence Length & 200 \\
LaMP & $B_{LaMP}$  & Minibatch Size & 200 \\
LaMP & $H_{LaMP}$  & LSTM Hidden Layer Size & 1500 \\
LaMP & $E_{LaMP}$  & Embedding Layer Size & 50 \\
LaMP & $D_{LaMP}$  & Dropout Ratio & 0.5 \\
\hline
\end{tabular}
\end{scriptsize}
\end{center}
\caption {Hyperparameter settings for the various models. The hyperparameter settings
for the \SCL variant are identical to the \PIL model.}
\label{tab:hyper}
\end{table}

\begin{table*}[tb]
\begin{center}
\begin{scriptsize}
\begin{tabular}{| c | c | c | c | c | c |}
\hline
Model & Accuracy (\%) & Precision (\%) & Recall (\%) & F1 & AUC \\
\hline
\hline
$\textsc{CPoLS-LSTM-LR}$ $(L=1, C=0, T=60K)$ & 98.8725 & 98.9990 & 99.7212 & 0.9936 & 0.9985 \\
$\textsc{CPoLS-LSTM-NN}$ $(L=1, C=1, T=60K)$ & 98.1997 & 98.2232 & 99.7492 & 0.9898 & 0.9937 \\
$\textsc{CPoLS-LSTM-DNN}$ $(L=1, C=2, T=60K)$ & 98.1966 & 98.1135 & 99.8615 & 0.9898 & 0.9964 \\
$\textsc{CPoLS-LSTM-LR}$ $(L=2, C=0, T=60K)$ & 99.0399 & 99.3888 & 99.5160 & 0.9945 & 0.9975 \\
$\textsc{CPoLS-LSTM-NN}$ $(L=2, C=1, T=60K)$ & 98.9708 & 99.1626 & 99.6668 & 0.9941 & 0.9984 \\
$\textsc{CPoLS-LSTM-DNN}$ $(L=2, C=2, T=60K)$ & 98.8771 & 99.1909 & 99.5301 & 0.9936 & 0.9981 \\

\hline
$\textsc{CPoLS-BiLSTM-LR}$ $(L=1, C=0, T=60K)$ & 98.5683 & 98.5394 & 99.8457 & 0.9919 & 0.9967 \\
$\textsc{CPoLS-BiLSTM-NN}$ $(L=1, C=1, T=60K)$ & 98.6928 & 98.7318 & 99.7896 & 0.9926 & 0.9956 \\
$\textsc{CPoLS-BiLSTM-DNN}$ $(L=1, C=2, T=60K)$ & 98.7035 & 98.8149 & 99.7159 & 0.9926 & 0.9969 \\
$\textsc{CPoLS-BiLSTM-LR}$ $(L=2, C=0, T=60K)$ & 98.7988 & 98.8773 & 99.7615 & 0.9932 & 0.9982 \\
$\textsc{CPoLS-BiLSTM-NN}$ $(L=2, C=1, T=60K)$ & 99.1398 & 99.2981 & 99.7229 & 0.9951 & 0.9986 \\
$\textsc{CPoLS-BiLSTM-DNN}$ $(L=2, C=2, T=60K)$ & 97.5530 & 97.4753 & 99.7913 & 0.9862 & 0.9969 \\

\hline
$\textsc{LaMP-LSTM-LR}$ $(L=1, C=0, T=200)$ & 95.9861 & 96.6608 & 98.8321 & 0.9773 & 0.9766 \\
$\textsc{LaMP-LSTM-NN}$ $(L=1, C=1, T=200)$ & 97.0138 & 96.9490 & 99.7295 & 0.9832 & 0.9892 \\
$\textsc{LaMP-LSTM-DNN}$ $(L=1, C=2, T=200)$ & 96.3953 & 96.4409 & 99.5592 & 0.9798 & 0.9873 \\
$\textsc{LaMP-LSTM-LR}$ $(L=2, C=0, T=200)$ & 87.5983 & 87.5983 & 100.0000 & 0.9339 & 0.5000 \\
$\textsc{LaMP-LSTM-NN}$ $(L=2, C=1, T=200)$ & 94.1814 & 96.0273 & 97.3866 & 0.9670 & 0.9500 \\
$\textsc{LaMP-LSTM-DNN}$ $(L=2, C=2, T=200)$ & 96.1169 & 97.8491 & 97.7151 & 0.9778 & 0.9748 \\

\hline
$\textsc{SdA-LR}$ $(T=2000)$ & 87.6020 & 87.6020 & 100.0000 & 0.9339 & 0.5012 \\
$\textsc{KOL-CNN}$ $(T=200)$ & 97.0753 & 97.4956 & 99.2097 & 0.9835 & 0.9853 \\
$\textsc{KOL-CNN}$ $(T=1000)$ & 96.7446 & 96.8356 & 99.5363 & 0.9817 & 0.9851 \\
$\textsc{LR - Trigram}$ $(T=60K)$ & 97.5975 & 97.9394 & 99.3477 & 0.9864 & 0.9237 \\
$\textsc{SVM - Trigram}$ $(T=60K)$ & 97.5560 & 97.7683 & 99.4810 & 0.9862 & 0.9185 \\

$\textsc{NB - Trigram}$ $(T=60K)$ & 97.5560 & 97.7683 & 99.4810 & 0.9862 & 0.9185 \\

\hline
\end{tabular}
\end{scriptsize}
\end{center}
\caption {Performance of the various models which were evaluated for this study.}
\label{tab:model-perf}
\end{table*}

\noindent\textbf{CPoLS Models:}
We first evaluate the performance of the \PIL model.
Their common performance metrics, along with the metrics of all the other models,
are summarized in Table~\ref{tab:model-perf}.  These performance metrics include the accuracy, precision, recall,
F1 score, and the area under the receiver operating characteristic
(ROC) curve (AUC). The table indicates
that, in general, most of the models perform reasonably well, although some models clearly outperform others. 

\begin{figure}[tb]
\centering
  \includegraphics[trim = 0.0in 0.0in 0.0in 0.0in,clip,width=1.0\columnwidth]{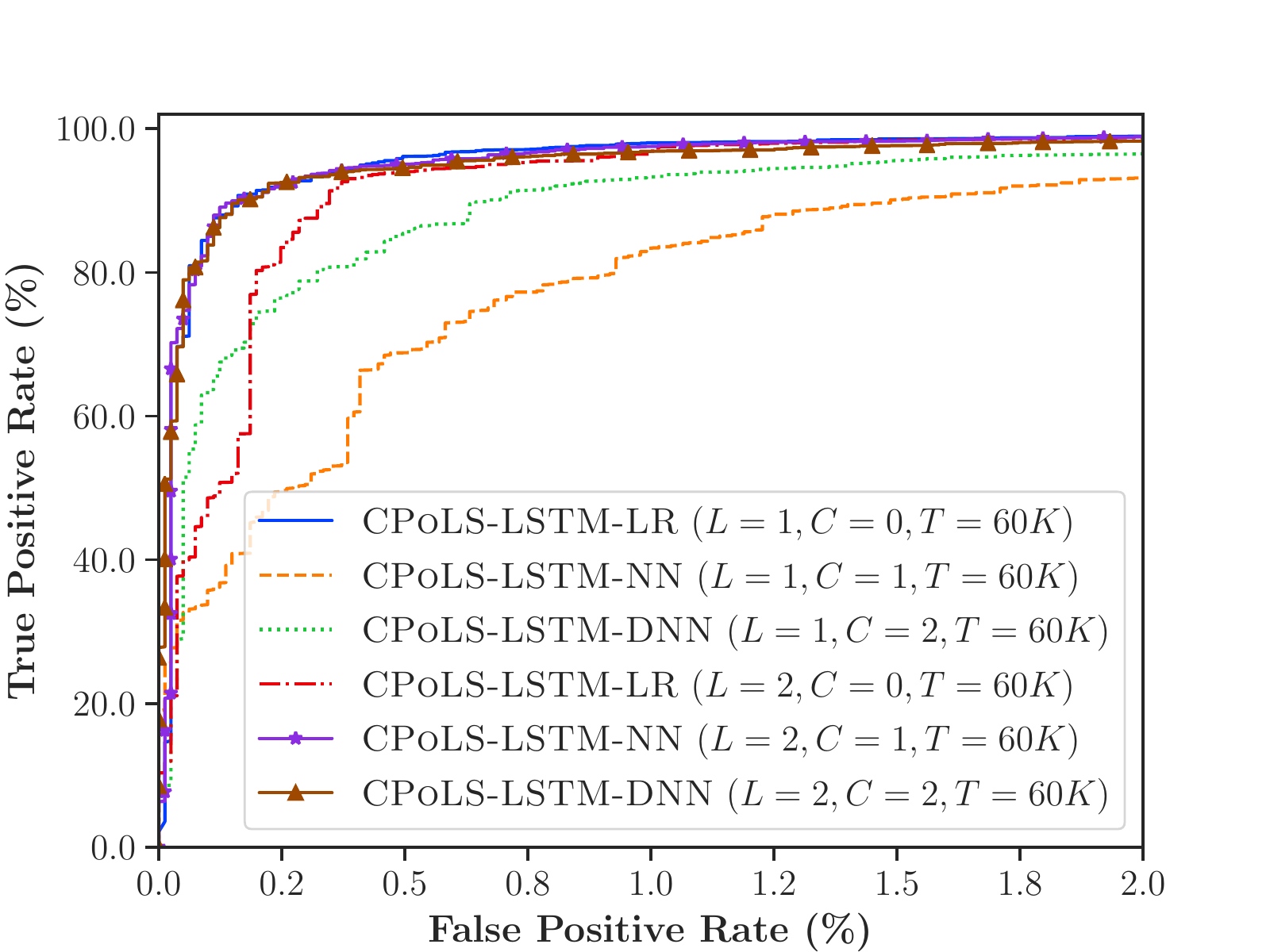}
  \caption{ROC curves for different JavaScript CPoLS models zoomed into a maximum FPR = 2\%.}
  \label{fig:JS-CPoLS-2}
\end{figure}

The ROC curves, which vary the FPR from 0\% to 2\%, for
the
CPoLS model with several different combinations of LSTM stacked layers $L_{CPoLS}$ and classifier hidden layers $C_{CPoLS}$, are depicted in
Figure~\ref{fig:JS-CPoLS-2}.
Even with the truncated JavaScript file sequences, all of the models
approximate an ideal classifier.
Above a false positive rate (FPR) of 0.15\%, the best performing \SCL model utilizes a single LSTM layer and single classifier hidden layer, $L_{CPoLS}=1, C_{CPoLS}=1$.
This result has several benefits.
Since the model has a fixed size, increasing the number of layers can often lead to overfitting the learned parameters in the model, leading to performance degradation on model evaluation.
Single layers also help limit the number of parameters of the \PIL model and make it faster and more compact for deployment at scale.
\begin{figure}[tb]
\centering
  \caption{ROC curves for different JavaScript CPoLS models for a maximum FPR = 100\%.}
  \label{fig:JS-CPoLS-100}
\end{figure}

\begin{figure}[tb]
\centering
  \includegraphics[trim = 0.0in 0.0in 0.0in 0.0in,clip,width=1.0\columnwidth]{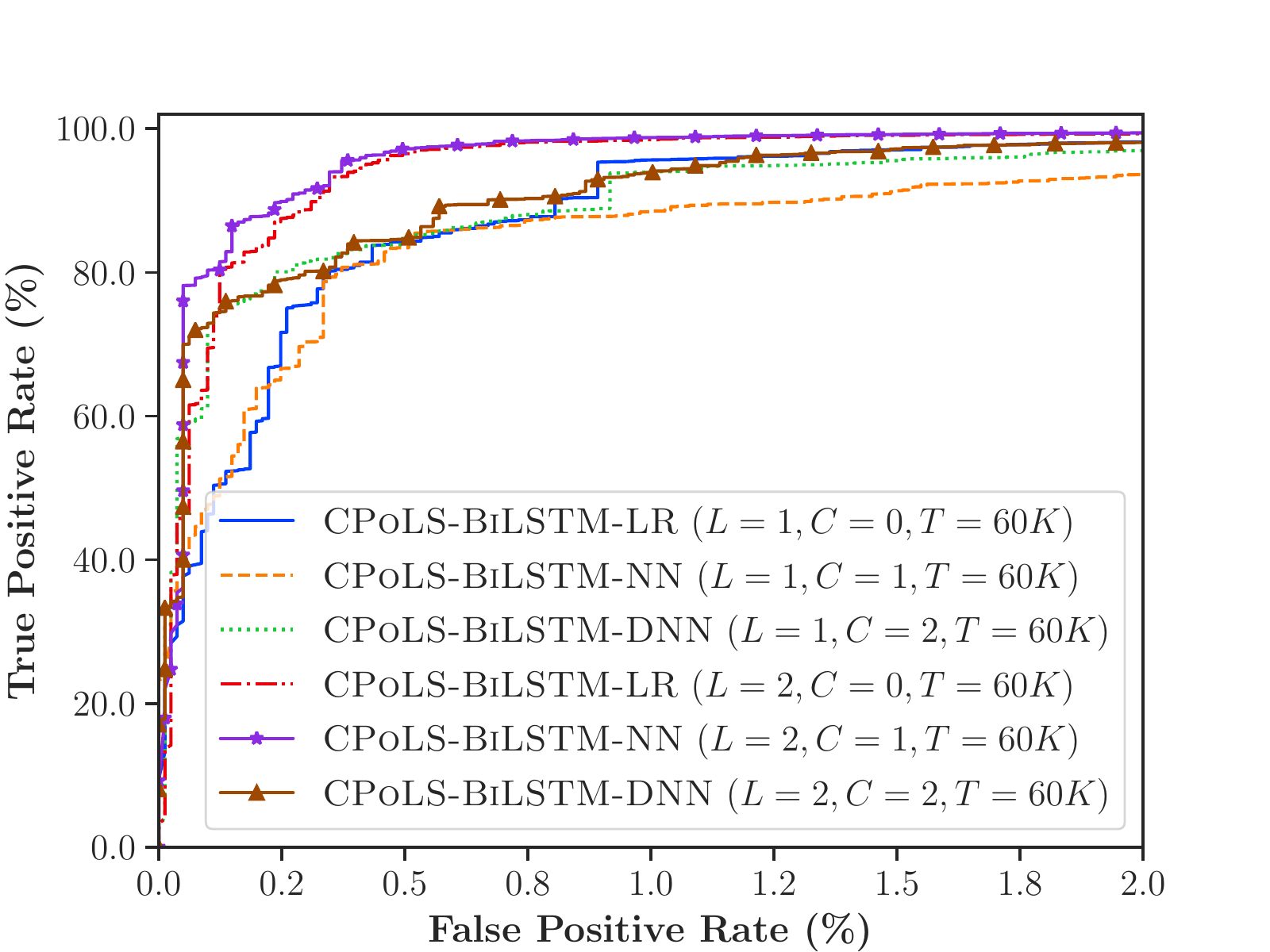}
  \caption{ROC curves for different JavaScript CPoLS-BiLSTM models zoomed into a maximum FPR = 2\%.}
  \label{fig:JS-CPoLS-BILSTM-2}
\end{figure}

We also evaluated the CPoLS architecture using the BiLSTMs.
The CPoLS-BiLSTM results are provided in Figure~\ref{fig:JS-CPoLS-BILSTM-2}.


%

\begin{figure}[tb]
\centering
  \includegraphics[trim = 0.0in 0.0in 0.0in 0.0in,clip,width=1.0\columnwidth]{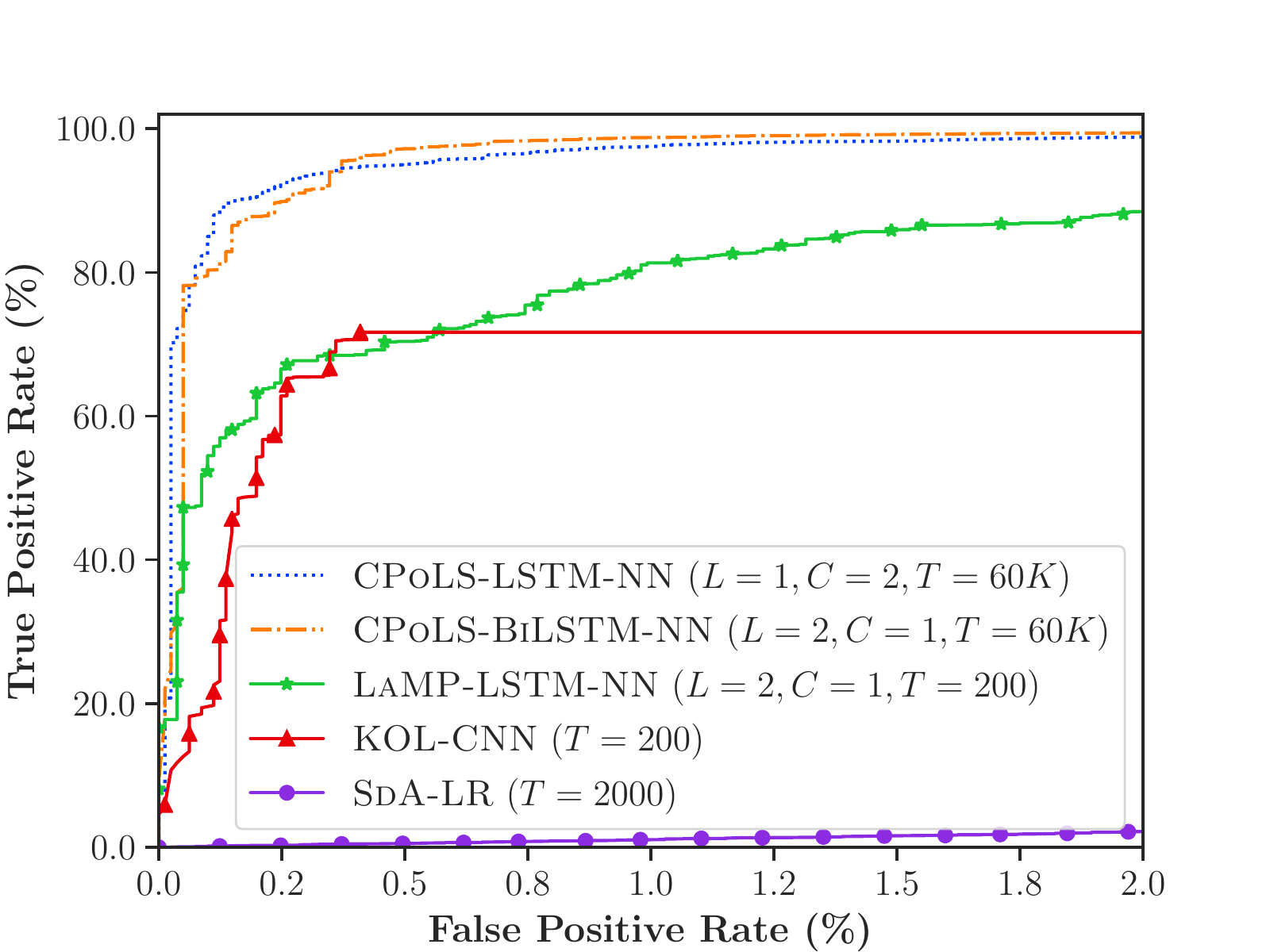}
  \caption{ROC curves for different JavaScript models zoomed into a maximum FPR = 2\%.}
  \label{fig:JS-Final-pdf}
\end{figure}

\noindent\textbf{Baselines:}
We now compare the performance results of the best performing \PIL and \SCL models to a number of baseline systems
summarized in 
Table~\ref{tab:model-perf}. 
The ROC curves for all of these models
are presented in Figure~\ref{fig:JS-Final-pdf}.

The \MPL model originally proposed in~\cite{BenMalware} for Windows PE
files is evaluated for this new task of detecting malicious JavaScript. Table~\ref{tab:model-perf} indicates that we
evaluated six variants  of the \MPL architecture in \Sys. Similarly, we implemented the sequential CNN model
proposed in~\cite{Kolosnjaji}, and denoted as KOL-CNN, which
is adapted
for the new task of detecting malicious JavaScript. Like~\cite{BenMalware}, this sequential KOL-CNN
model was proposed to detect Windows PE files. We also re-implemented the SDA-LR model~\cite{sdalr_wang} which uses autoencoders to detect malicious JavaScript.
We also compare against trigrams of byte using logistic regression (LR-Trigram) and a support vector machine (SVM-Trigram) as proposed in~\cite{Shah2016}.
Naive Bayes with trigrams is also considered.

None of these models are designed to process very long sequences. In fact, we tried to implement the \MPL models with length $T$ = 1000 JavaScript bytes, but all
those experiments
generated out of memory exceptions. We were able to process KOL-CNN with length $T$ = 1000 sequences. We were also able to process length  $T$ = 2000 sequences
with SDA-LR.

As indicated in Figure~\ref{fig:JS-Final-pdf}, none of these baseline models outperformed our models on the JavaScript data
files. In particular, the SDA-LR model predicted that all the JavaScript files in the test set were malicious for a number of variants that we explored.

\section{Related Work}
\label{sec:related}
%
\noindent\textbf{JavaScript:}
Maiorca \textit{et al.}~\cite{Maiorca15} propose a static analysis-based system to detect malicious
PDF files which use features constructed from both the content of the PDF, including JavaScript, as well as its structure. Once
these features are extracted, the authors use a boosted decision tree trained with the AdaBoost algorithm to detect malicious PDFs.
%
%

Cova \textit{et al.}~\cite{Cova2010} use the approach of anomaly detection for detecting malicious JavaScript code.
They learn a model for representing normal (benign) JavaScript code, and then use it during the detection of anomalous code.
They also present the learning of specific features that helps characterize intrinsic events of a drive-by download.

Hallaraker and Vigna~\cite{Hallaraker2005} present an auditing system in Mozilla  for JavaScript interpreters.
They provide logging and monitoring on downloaded JavaScript, which can be integrated with
intrusion detection systems for malicious behavior detection.

In~\cite{Likarish2009}, Likarish \textit{et al.} classify obfuscated malicious JavaScript using several different types of classifiers including Naive Bayes, an Alternating Decision Tree (ADTree), a Support Vector Machine (SVM) with using the Radial Basis Function (RBF) kernel, and the rule-based Ripper algorithm. In their static analysis-based study, the SVM performed best based on tokenized unigrams and bigrams chosen by feature selection.

A PDF classifier proposed by Laskov and \v{S}rndi\'{c}~\cite{Laskov2011} uses a one-class SVM to
detect malicious PDFs which contain JavaScript code. Laskov's system is based solely on static
analysis. The features are derived from lexical analysis of JavaScript code extracted from the PDF
files in their dataset.

Zozzle~\cite{Zozzle} proposes a mostly static approach extracting contexts from the original JavaScript file. The system parses these contexts to recover the abstract syntax trees (ASTs).
A Naive Bayes classifier is then trained on the features extracted from the variables and keyword found in the ASTs.

Corona \textit{et al.}~\cite{Corona2014}, propose Lux0R, a system to select API references for the detection of malicious JavaScript in PDF documents. These
references include JavaScript APIs as well as functions, methods, keywords, and constants. The authors propose a discriminant analysis feature selection method.
The features are then classified with an SVM, a Decision Tree and a Random Forest model.
Like ScriptNet, Lux0R performs both static and dynamic analysis. However, they do not use deep
learning and require the extraction of the JavaScript API references.

Wang \textit{et al.}~\cite{Wang2016} use deep learning models in combination with sparse random projections, and logistic regression.
They also present feature extraction from JavaScript code using auto-encoders.
While they use deep learning models, the feature extraction and model architectures limit the information extractability from JavaScript code.

Like our work, several authors have proposed different types of static JavaScript classifiers which just analyzes the raw script content.
Shah~\cite{Shah2016} propose using a statistical n-gram language model
to detect malicious JavaScript. Our proposed system uses an LSTM neural model for the language model instead of the n-gram model proposed by Shah~\cite{Shah2016}.
Other papers which investigate the detection of malicious JavaScript include~\cite{Liu2014,Schutt2012,Wang2013,Xu2012,Xu2013}.

\noindent\textbf{Other File Types:}
While more research has been devoted to detecting malicious JavaScript, partly because of
its inclusion in malicious PDFs, only a few previous studies have considered malicious VBScript.
In~\cite{Kim2006}, a conceptual graph is first computed for VBScript files, and new malware
is detected by identifying graphs which are similar to those of known malicious VBScript files. The
method is based on static analysis of the VBScripts.
%
Wael \textit{et al.}~\cite{Wael2017} propose a number of different classifiers to detect malicious VBScript including Logistic Regression, a Support Vector Machine with an RBF kernel, a Random Forest, a Multilayer Perceptron, and a Decision Table. The features are created based on static analysis. The best performing classifier in their study is the SVM. In~\cite{Zhao2010}, Zhao and Chen detect malicious applets, JavaScript and VBScript based on a method which models immunoglobulin secretion.

A number of deep learning models have been proposed for detecting malicious PE files including~\cite{BenMalware,Dahl2013,Huang2016,Kolosnjaji,PascanuMalware}.
In particular, a character-level CNN has been proposed for detecting malicious PE files~\cite{BenMalware} and Powershell script files~\cite{Hendler2018}.
Raff \textit{et al.}~\cite{Raff2017} discuss a model which is similar to CPoLS but noted it did not work for PE files. They did not provide any results
for their model.

\section{Conclusions}
\label{sec:conc}
Malicious JavaScript detection is an important
problem facing anti-virus companies. Failure to
detect a malicious JavaScript file may result in a
successful spearphishing, ransomware, or drive-by
download attack. Neural language models have shown promising
results in the detection of malicious executable files.
Similarly, we show that these types of models
can also detect malicious JavaScript
files, in the proposed \Sys system, with very high true
positive rates at extremely low false positive rates.

The performance results confirm that the \PIL model
using CNN 
and LSTM neural layers
is able to learn and generate representations of byte sequences
in the JavaScript files.
In particular, the \PIL JavaScript malware script classification model using a
single LSTM layer and a shallow neural network layer offers the best results.
Therefore, the vector representations generated by these models capture important
sequential information from the JavaScript files.
\Sys extracts and uses this information to predict the malicious intent of these files.

\sloppy
\section*{Acknowledgement}
The authors thank Marc Marino, Jugal Parikh, Daewoo Chong, Mikael Figueroa and Arun Gururajan for providing the data and helpful discussions.

\vskip 0.2in
\bibliographystyle{IEEEtranS}
\balance
\bibliography{usenix2019}

\begin{thebibliography}{10}

\bibitem{virustotal}
Virustotal.

\bibitem{Tensorflow}
{\sc Abadi, M., Agarwal, A., Barham, P., Brevdo, E., Chen, Z., Citro, C.,
  Corrado, G.~S., Davis, A., Dean, J., Devin, M., Ghemawat, S., Goodfellow, I.,
  Harp, A., Irving, G., Isard, M., Jia, Y., Jozefowicz, R., Kaiser, L., Kudlur,
  M., Levenberg, J., Man\'{e}, D., Monga, R., Moore, S., Murray, D., Olah, C.,
  Schuster, M., Shlens, J., Steiner, B., Sutskever, I., Talwar, K., Tucker, P.,
  Vanhoucke, V., Vasudevan, V., Vi\'{e}gas, F., Vinyals, O., Warden, P.,
  Wattenberg, M., Wicke, M., Yu, Y., and Zheng, X.}
\newblock {TensorFlow}: Large-scale machine learning on heterogeneous systems,
  2015.
\newblock Software available from tensorflow.org.

\bibitem{BenMalware}
{\sc Athiwaratkun, B., and Stokes, J.~W.}
\newblock Malware classification with lstm and gru language models and a
  character-level cnn.
\newblock In {\em 2017 IEEE International Conference on Acoustics, Speech and
  Signal Processing (ICASSP)\/} (March 2017), pp.~2482--2486.

\bibitem{BengioLongTerm1994}
{\sc Bengio, Y., Simard, P., and Frasconi, P.}
\newblock Learning long-term dependencies with gradient descent is difficult.
\newblock {\em IEEE Transactions on Neural Networks 5}, 2 (Mar 1994), 157--166.

\bibitem{Cho2014}
{\sc Cho, K., van Merrienboer, B., G{\"{u}}l{\c{c}}ehre, {\c{C}}., Bougares,
  F., Schwenk, H., and Bengio, Y.}
\newblock Learning phrase representations using {RNN} encoder-decoder for
  statistical machine translation.
\newblock {\em CoRR abs/1406.1078\/} (2014).

\bibitem{keras}
{\sc Chollet, F., et~al.}
\newblock Keras.
\newblock \url{https://github.com/fchollet/keras}, 2015.

\bibitem{Corona2014}
{\sc Corona, I., Maiorca, D., Ariu, D., and Giacinto, G.}
\newblock Lux0r: Detection of malicious pdf-embedded javascript code through
  discriminant analysis of api references.
\newblock In {\em Proceedings of the 2014 Workshop on Artificial Intelligent
  and Security Workshop\/} (New York, NY, USA, 2014), AISec '14, ACM,
  pp.~47--57.

\bibitem{mmpc_2016_js_ransomware}
{\sc Corporation, M.}
\newblock {Don’t let this Black Friday/Cyber Monday spam deliver Locky
  ransomware to you}, 2016.

\bibitem{Cova2010}
{\sc Cova, M., Kruegel, C., and Vigna, G.}
\newblock Detection and analysis of drive-by-download attacks and malicious
  javascript code.
\newblock In {\em Proceedings of the 19th International Conference on World
  Wide Web\/} (New York, NY, USA, 2010), WWW '10, ACM, pp.~281--290.

\bibitem{crnBreach}
{\sc CRN}.
\newblock Pentagon data breach shows growing sophistication of phishing
  attacks.

\bibitem{Dahl2013}
{\sc Dahl, G.~E., Stokes, J.~W., Deng, L., and Yu, D.}
\newblock Large-scale malware classification using random projections and
  neural networks.
\newblock In {\em Proceedings of the IEEE International Conference on
  Acoustics, Speech and Signal Processing (ICASSP)\/} (2013).

\bibitem{Gandotra2014}
{\sc Gandotra, E., Bansal, D., and Sofat, S.}
\newblock Malware analysis and classification: A survey.
\newblock 55--64.

\bibitem{Gehring2016}
{\sc Gehring, J., Auli, M., Grangier, D., and Dauphin, Y.~N.}
\newblock A convolutional encoder model for neural machine translation.
\newblock {\em CoRR abs/1611.02344\/} (2016).

\bibitem{Gehring2017}
{\sc Gehring, J., Auli, M., Grangier, D., Yarats, D., and Dauphin, Y.~N.}
\newblock Convolutional sequence to sequence learning.
\newblock {\em CoRR abs/1705.03122\/} (2017).

\bibitem{GersJj1999}
{\sc Gers, F.~A., Schmidhuber, J., and Cummins, F.~A.}
\newblock Learning to forget: Continual prediction with {LSTM}.
\newblock {\em Neural Computation 12}, 10 (2000), 2451--2471.

\bibitem{glorot10a}
{\sc Glorot, X., and Bengio, Y.}
\newblock Understanding the difficulty of training deep feedforward neural
  networks.
\newblock In {\em Proceedings of the Thirteenth International Conference on
  Artificial Intelligence and Statistics\/} (Chia Laguna Resort, Sardinia,
  Italy, 13--15 May 2010), Y.~W. Teh and M.~Titterington, Eds., vol.~9 of {\em
  Proceedings of Machine Learning Research}, PMLR, pp.~249--256.

\bibitem{Graves2013b}
{\sc Graves, A., Jaitly, N., and Mohamed, A.-r.}
\newblock {Hybrid speech recognition with Deep Bidirectional LSTM}.
\newblock In {\em 2013 IEEE Workshop on Automatic Speech Recognition and
  Understanding\/} (dec 2013), IEEE, pp.~273--278.

\bibitem{GravesSpeech}
{\sc Graves, A., r.~Mohamed, A., and Hinton, G.}
\newblock Speech recognition with deep recurrent neural networks.
\newblock In {\em 2013 IEEE International Conference on Acoustics, Speech and
  Signal Processing\/} (May 2013), pp.~6645--6649.

\bibitem{Grosse2017a}
{\sc Grosse, K., Papernot, N., Manoharan, P., Backes, M., and McDaniel, P.}
\newblock Adversarial perturbations against deep neural networks for malware
  classification.
\newblock In {\em Proceedings of the European Symposium on Research in Computer
  Security (ESORICS)\/} (2017).

\bibitem{Hallaraker2005}
{\sc Hallaraker, O., and Vigna, G.}
\newblock Detecting malicious javascript code in mozilla.
\newblock In {\em 10th IEEE International Conference on Engineering of Complex
  Computer Systems (ICECCS'05)\/} (June 2005), pp.~85--94.

\bibitem{Hendler2018}
{\sc {Hendler}, D., {Kels}, S., and {Rubin}, A.}
\newblock {Detecting Malicious PowerShell Commands using Deep Neural Networks}.
\newblock {\em ArXiv e-prints\/} (Apr. 2018).

\bibitem{Hochreiter1998}
{\sc Hochreiter, S.}
\newblock The vanishing gradient problem during learning recurrent neural nets
  and problem solutions.
\newblock {\em Int. J. Uncertain. Fuzziness Knowl.-Based Syst. 6}, 2 (Apr.
  1998), 107--116.

\bibitem{Hochreiter1997}
{\sc Hochreiter, S., and Schmidhuber, J.}
\newblock {Long short-term memory}.
\newblock {\em Neural Computation 9}, 8 (1997), 1--32.

\bibitem{HuAdversarialMalwareGan}
{\sc Hu, W., and Tan, Y.}
\newblock Generating adversarial malware examples for black-box attacks based
  on gan.
\newblock {\em arXiv preprint 1702.05983\/} (2017).

\bibitem{Huang2016}
{\sc Huang, W., and Stokes, J.~W.}
\newblock Mtnet: A multi-task neural network for dynamic malware classfication.
\newblock In {\em Proceedings of Detection of Intrusions and Malware, and
  Vulnerability Assessment (DIMVA)\/} (2016), pp.~399--418.

\bibitem{Kim2006}
{\sc Kim, S., Choi, C., Choi, J., Kim, P., and Kim, H.}
\newblock A method for efficient malicious code detection based on conceptual
  similarity.
\newblock In {\em International Conference on Computational Science and Its
  Applications (ICCSA)\/} (2006), vol.~3983, pp.~567--576.

\bibitem{Kingma2014}
{\sc Kingma, D.~P., and Ba, J.}
\newblock {Adam: A Method for Stochastic Optimization}.

\bibitem{Kolosnjaji}
{\sc Kolosnjaji, B., Zarras, A., Webster, G., and Eckert, C.}
\newblock Deep learning for classification of malware system call sequences.
\newblock In {\em Australasian Joint Conference on Artificial Intelligence\/}
  (2016), Springer International Publishing, pp.~137--149.

\bibitem{krizhevsky2012imagenet}
{\sc Krizhevsky, A., Sutskever, I., and Hinton, G.~E.}
\newblock Imagenet classification with deep convolutional neural networks.
\newblock In {\em Advances in neural information processing systems\/} (2012),
  pp.~1097--1105.

\bibitem{Laskov2011}
{\sc Laskov, P., and \v{S}rndi\'{c}, N.}
\newblock Static detection of malicious javascript-bearing pdf documents.
\newblock In {\em Proceedings of the 27th Annual Computer Security Applications
  Conference\/} (New York, NY, USA, 2011), ACSAC '11, ACM, pp.~373--382.

\bibitem{LeCun1995ConvolutionalNF}
{\sc LeCun, Y., and Bengio, Y.}
\newblock Convolutional networks for images speech and time series.

\bibitem{Likarish2009}
{\sc Likarish, P., Jung, E., and Jo, I.}
\newblock {Obfuscated malicious javascript detection using classification
  techniques}.
\newblock In {\em 2009 4th International Conference on Malicious and Unwanted
  Software (MALWARE)\/} (oct 2009), IEEE, pp.~47--54.

\bibitem{Liu2014}
{\sc Liu, D., Wang, H., and Stavrou, A.}
\newblock Detecting malicious javascript in pdf through document
  instrumentation.
\newblock In {\em 2014 44th Annual IEEE/IFIP International Conference on
  Dependable Systems and Networks\/} (June 2014), pp.~100--111.

\bibitem{Maiorca15}
{\sc Maiorca, D., Ariu, D., Corona, I., and Giacinto, G.}
\newblock A structural and content-based approach for a precise and robust
  detection of malicious pdf files.
\newblock In {\em Proceedings of the International Conference on Information
  Systems Security and Privacy (ICISSP)\/} (2015).

\bibitem{JS}
{\sc Mozilla}.
\newblock {JavaScript}.

\bibitem{ohrimenko2016oblivious}
{\sc Ohrimenko, O., Schuster, F., Fournet, C., Mehta, A., Nowozin, S., Vaswani,
  K., and Costa, M.}
\newblock Oblivious multi-party machine learning on trusted processors.
\newblock In {\em USENIX Security Symposium\/} (2016), pp.~619--636.

\bibitem{papernot2015limitations}
{\sc Papernot, N., McDaniel, P., Jha, S., Fredrikson, M., Celik, Z.~B., and
  Swami, A.}
\newblock The limitations of deep learning in adversarial settings.
\newblock {\em Proceedings of the 1st IEEE European Symposium on Security and
  Privacy\/} (2015).

\bibitem{PapernotRnn2016}
{\sc Papernot, N., McDaniel, P., Swami, A., and Harang, R.}
\newblock Crafting adversarial input sequences for recurrent neural networks.
\newblock In {\em Proceedings of the Military Communications Conference
  (MILCOM)\/} (2016).

\bibitem{PascanuMalware}
{\sc Pascanu, R., Stokes, J.~W., Sanossian, H., Marinescu, M., and Thomas, A.}
\newblock Malware classification with recurrent networks.
\newblock In {\em 2015 IEEE International Conference on Acoustics, Speech and
  Signal Processing (ICASSP)\/} (April 2015), pp.~1916--1920.

\bibitem{Raff2017}
{\sc {Raff}, E., {Barker}, J., {Sylvester}, J., {Brandon}, R., {Catanzaro}, B.,
  and {Nicholas}, C.}
\newblock {Malware Detection by Eating a Whole EXE}.
\newblock {\em ArXiv e-prints\/} (2017).

\bibitem{russakovsky2015imagenet}
{\sc Russakovsky, O., Deng, J., Su, H., Krause, J., Satheesh, S., Ma, S.,
  Huang, Z., Karpathy, A., Khosla, A., Bernstein, M., et~al.}
\newblock Imagenet large scale visual recognition challenge.
\newblock {\em International Journal of Computer Vision 115}, 3 (2015),
  211--252.

\bibitem{Schutt2012}
{\sc Sch\"{u}tt, K., Kloft, M., Bikadorov, A., and Rieck, K.}
\newblock Early detection of malicious behavior in javascript code.
\newblock In {\em Proceedings of the 5th ACM Workshop on Security and
  Artificial Intelligence\/} (New York, NY, USA, 2012), AISec '12, ACM,
  pp.~15--24.

\bibitem{Shah2016}
{\sc Shah, A.}
\newblock {Malicious JavaScript Detection using Statistical Language Model}.
\newblock {\em Master's Projects\/} (2016), 70.

\bibitem{verizonBreach}
{\sc Snell, E.}
\newblock Verizon finds phishing attacks, malware top data breach causes.

\bibitem{Stokes2017Adversarial}
{\sc {Stokes}, J.~W., {Wang}, D., {Marinescu}, M., {Marino}, M., and {Bussone},
  B.}
\newblock {Attack and Defense of Dynamic Analysis-Based, Adversarial Neural
  Malware Classification Models}.
\newblock {\em ArXiv e-prints\/} (Dec. 2017).

\bibitem{SutskeverSeq2Seq}
{\sc Sutskever, I., Vinyals, O., and Le, Q.~V.}
\newblock Sequence to sequence learning with neural networks.
\newblock In {\em Advances in Neural Information Processing Systems 27},
  Z.~Ghahramani, M.~Welling, C.~Cortes, N.~D. Lawrence, and K.~Q. Weinberger,
  Eds. Curran Associates, Inc., 2014, pp.~3104--3112.

\bibitem{Wael2017}
{\sc Wael, D., Shosha, A., and Sayed, S.~G.}
\newblock Malicious vbscript detection algorithm based on data-mining
  techniques.
\newblock In {\em 2017 Intl Conf on Advanced Control Circuits Systems (ACCS)
  Systems 2017 Intl Conf on New Paradigms in Electronics Information Technology
  (PEIT)\/} (Nov 2017), pp.~112--116.

\bibitem{Wang2013}
{\sc Wang, W.-H., Lv, Y.-J., Chen, H.-B., and Fang, Z.-L.}
\newblock A static malicious javascript detection using svm.
\newblock In {\em Proceedings of the 2nd International Conference on Computer
  Science and Electronics Engineering\/} (2013).

\bibitem{sdalr_wang}
{\sc Wang, Y., Cai, W.-d., and Wei, P.-c.}
\newblock A deep learning approach for detecting malicious javascript code.
\newblock {\em Security and Communication Networks 9}, 11, 1520--1534.

\bibitem{Wang2016}
{\sc Wang, Y., dong Cai, W., and cheng Wei, P.}
\newblock A deep learning approach for detecting malicious javascript code.
\newblock {\em Proceedings of Security and Communication Networks 11}, 9
  (2016), 1520--1534.

\bibitem{Werbos1990}
{\sc Werbos, P.~J.}
\newblock {Backpropagation Through Time: What It Does and How to Do It}.
\newblock {\em Proceedings of the IEEE 78}, 10 (1990), 1550--1560.

\bibitem{adversarial_xiao_li_song}
{\sc Xiao, C., Li, B., Zhu, J., He, W., Liu, M., and Song, D.}
\newblock Generating adversarial examples with adversarial networks.
\newblock {\em CoRR abs/1801.02610\/} (2018).

\bibitem{adversarial_xiao_zhu_li_song}
{\sc Xiao, C., Zhu, J.-Y., Li, B., He, W., Liu, M., and Song, D.}
\newblock Spatially transformed adversarial examples.
\newblock In {\em International Conference on Learning Representations\/}
  (2018).

\bibitem{evans_adversarial}
{\sc Xu, W., Evans, D., and Qi, Y.}
\newblock Feature squeezing: Detecting adversarial examples in deep neural
  networks.
\newblock {\em CoRR abs/1704.01155\/} (2017).

\bibitem{Xu2012}
{\sc Xu, W., Zhang, F., and Zhu, S.}
\newblock The power of obfuscation techniques in malicious javascript code: A
  measurement study.
\newblock In {\em 2012 7th International Conference on Malicious and Unwanted
  Software\/} (Oct 2012), pp.~9--16.

\bibitem{Xu2013}
{\sc Xu, W., Zhang, F., and Zhu, S.}
\newblock Jstill: Mostly static detection of obfuscated malicious javascript
  code.
\newblock In {\em Proceedings of the Third ACM Conference on Data and
  Application Security and Privacy\/} (New York, NY, USA, 2013), CODASPY '13,
  ACM, pp.~117--128.

\bibitem{Zhao2010}
{\sc Zhao, H., and Chen, W.}
\newblock A web page malicious script detection method inspired by the process
  of immunoglobulin secretion.
\newblock In {\em 2010 International Symposium on Intelligence Information
  Processing and Trusted Computing\/} (Oct 2010), pp.~241--245.

\end{thebibliography}


\begin{thebibliography}{10}
\providecommand{\url}[1]{#1}
\csname url@samestyle\endcsname
\providecommand{\newblock}{\relax}
\providecommand{\bibinfo}[2]{#2}
\providecommand{\BIBentrySTDinterwordspacing}{\spaceskip=0pt\relax}
\providecommand{\BIBentryALTinterwordstretchfactor}{4}
\providecommand{\BIBentryALTinterwordspacing}{\spaceskip=\fontdimen2\font plus
\BIBentryALTinterwordstretchfactor\fontdimen3\font minus
  \fontdimen4\font\relax}
\providecommand{\BIBforeignlanguage}[2]{{%
\expandafter\ifx\csname l@#1\endcsname\relax
\typeout{** WARNING: IEEEtranS.bst: No hyphenation pattern has been}%
\typeout{** loaded for the language `#1'. Using the pattern for}%
\typeout{** the default language instead.}%
\else
\language=\csname l@#1\endcsname
\fi
#2}}
\providecommand{\BIBdecl}{\relax}
\BIBdecl

\bibitem{Tensorflow}
\BIBentryALTinterwordspacing
M.~Abadi, A.~Agarwal, P.~Barham, E.~Brevdo, Z.~Chen, C.~Citro, G.~S. Corrado,
  A.~Davis, J.~Dean, M.~Devin, S.~Ghemawat, I.~Goodfellow, A.~Harp, G.~Irving,
  M.~Isard, Y.~Jia, R.~Jozefowicz, L.~Kaiser, M.~Kudlur, J.~Levenberg,
  D.~Man\'{e}, R.~Monga, S.~Moore, D.~Murray, C.~Olah, M.~Schuster, J.~Shlens,
  B.~Steiner, I.~Sutskever, K.~Talwar, P.~Tucker, V.~Vanhoucke, V.~Vasudevan,
  F.~Vi\'{e}gas, O.~Vinyals, P.~Warden, M.~Wattenberg, M.~Wicke, Y.~Yu, and
  X.~Zheng, ``{TensorFlow}: Large-scale machine learning on heterogeneous
  systems,'' 2015, software available from tensorflow.org. [Online]. Available:
  \url{http://tensorflow.org/}
\BIBentrySTDinterwordspacing

\bibitem{BenMalware}
B.~Athiwaratkun and J.~W. Stokes, ``Malware classification with lstm and gru
  language models and a character-level cnn,'' in \emph{2017 IEEE International
  Conference on Acoustics, Speech and Signal Processing (ICASSP)}, March 2017,
  pp. 2482--2486.

\bibitem{BengioLongTerm1994}
Y.~Bengio, P.~Simard, and P.~Frasconi, ``Learning long-term dependencies with
  gradient descent is difficult,'' \emph{IEEE Transactions on Neural Networks},
  vol.~5, no.~2, pp. 157--166, Mar 1994.

\bibitem{keras}
F.~Chollet \emph{et~al.}, ``Keras,'' \url{https://github.com/fchollet/keras},
  2015.

\bibitem{Corona2014}
I.~Corona, D.~Maiorca, D.~Ariu, and G.~Giacinto, ``Lux0r: Detection of
  malicious pdf-embedded javascript code through discriminant analysis of api
  references,'' in \emph{Proceedings of the 2014 Workshop on Artificial
  Intelligent and Security Workshop}, ser. AISec '14.\hskip 1em plus 0.5em
  minus 0.4em\relax New York, NY, USA: ACM, 2014, pp. 47--57.

\bibitem{Cova2010}
M.~Cova, C.~Kruegel, and G.~Vigna, ``Detection and analysis of
  drive-by-download attacks and malicious javascript code,'' in
  \emph{Proceedings of the 19th International Conference on World Wide Web},
  ser. WWW '10.\hskip 1em plus 0.5em minus 0.4em\relax New York, NY, USA: ACM,
  2010, pp. 281--290.

\bibitem{Zozzle}
C.~Curtsinger, B.~Livshits, B.~Zorn, and C.~Seifert, ``Zozzle: Fast and precise
  in-browser javascript malware detection,'' in \emph{Proceedings of Usenix
  Security}, 2011.

\bibitem{Dahl2013}
G.~E. Dahl, J.~W. Stokes, L.~Deng, and D.~Yu, ``Large-scale malware
  classification using random projections and neural networks,'' in
  \emph{Proceedings of the IEEE International Conference on Acoustics, Speech
  and Signal Processing (ICASSP)}, 2013.

\bibitem{Gandotra2014}
E.~Gandotra, D.~Bansal, and S.~Sofat, ``Malware analysis and classification: A
  survey,'' pp. 55--64, 2014.

\bibitem{Gehring2016}
\BIBentryALTinterwordspacing
J.~Gehring, M.~Auli, D.~Grangier, and Y.~N. Dauphin, ``A convolutional encoder
  model for neural machine translation,'' \emph{CoRR}, vol. abs/1611.02344,
  2016. [Online]. Available: \url{http://arxiv.org/abs/1611.02344}
\BIBentrySTDinterwordspacing

\bibitem{Gehring2017}
\BIBentryALTinterwordspacing
J.~Gehring, M.~Auli, D.~Grangier, D.~Yarats, and Y.~N. Dauphin, ``Convolutional
  sequence to sequence learning,'' \emph{CoRR}, vol. abs/1705.03122, 2017.
  [Online]. Available: \url{http://arxiv.org/abs/1705.03122}
\BIBentrySTDinterwordspacing

\bibitem{GersJj1999}
F.~A. Gers, J.~Schmidhuber, and F.~A. Cummins, ``Learning to forget: Continual
  prediction with {LSTM},'' \emph{Neural Computation}, vol.~12, no.~10, pp.
  2451--2471, 2000.

\bibitem{glorot10a}
\BIBentryALTinterwordspacing
X.~Glorot and Y.~Bengio, ``Understanding the difficulty of training deep
  feedforward neural networks,'' in \emph{Proceedings of the Thirteenth
  International Conference on Artificial Intelligence and Statistics}, ser.
  Proceedings of Machine Learning Research, Y.~W. Teh and M.~Titterington,
  Eds., vol.~9.\hskip 1em plus 0.5em minus 0.4em\relax Chia Laguna Resort,
  Sardinia, Italy: PMLR, 13--15 May 2010, pp. 249--256. [Online]. Available:
  \url{http://proceedings.mlr.press/v9/glorot10a.html}
\BIBentrySTDinterwordspacing

\bibitem{Hallaraker2005}
O.~Hallaraker and G.~Vigna, ``Detecting malicious javascript code in mozilla,''
  in \emph{10th IEEE International Conference on Engineering of Complex
  Computer Systems (ICECCS'05)}, June 2005, pp. 85--94.

\bibitem{Hendler2018}
D.~{Hendler}, S.~{Kels}, and A.~{Rubin}, ``{Detecting Malicious PowerShell
  Commands using Deep Neural Networks},'' \emph{ArXiv e-prints}, Apr. 2018.

\bibitem{Hochreiter1998}
\BIBentryALTinterwordspacing
S.~Hochreiter, ``The vanishing gradient problem during learning recurrent
  neural nets and problem solutions,'' \emph{Int. J. Uncertain. Fuzziness
  Knowl.-Based Syst.}, vol.~6, no.~2, pp. 107--116, Apr. 1998. [Online].
  Available: \url{http://dx.doi.org/10.1142/S0218488598000094}
\BIBentrySTDinterwordspacing

\bibitem{Hochreiter1997}
S.~Hochreiter and J.~Schmidhuber, ``{Long short-term memory},'' \emph{Neural
  Computation}, vol.~9, no.~8, pp. 1--32, 1997.

\bibitem{Huang2016}
W.~Huang and J.~W. Stokes, ``Mtnet: A multi-task neural network for dynamic
  malware classfication,'' in \emph{Proceedings of Detection of Intrusions and
  Malware, and Vulnerability Assessment (DIMVA)}, 2016, pp. 399--418.

\bibitem{Kim2006}
S.~Kim, C.~Choi, J.~Choi, P.~Kim, and H.~Kim, ``A method for efficient
  malicious code detection based on conceptual similarity,'' in
  \emph{International Conference on Computational Science and Its Applications
  (ICCSA)}, vol. 3983, 2006, pp. 567--576.

\bibitem{Kingma2014}
\BIBentryALTinterwordspacing
D.~P. Kingma and J.~Ba, ``{Adam: A Method for Stochastic Optimization},'' dec
  2014. [Online]. Available: \url{http://arxiv.org/abs/1412.6980}
\BIBentrySTDinterwordspacing

\bibitem{Kolosnjaji}
B.~Kolosnjaji, A.~Zarras, G.~Webster, and C.~Eckert, ``Deep learning for
  classification of malware system call sequences,'' in \emph{Australasian
  Joint Conference on Artificial Intelligence}.\hskip 1em plus 0.5em minus
  0.4em\relax Springer International Publishing, 2016, pp. 137--149.

\bibitem{krizhevsky2012imagenet}
A.~Krizhevsky, I.~Sutskever, and G.~E. Hinton, ``Imagenet classification with
  deep convolutional neural networks,'' in \emph{Advances in neural information
  processing systems}, 2012, pp. 1097--1105.

\bibitem{Laskov2011}
P.~Laskov and N.~\v{S}rndi\'{c}, ``Static detection of malicious
  javascript-bearing pdf documents,'' in \emph{Proceedings of the 27th Annual
  Computer Security Applications Conference}, ser. ACSAC '11.\hskip 1em plus
  0.5em minus 0.4em\relax New York, NY, USA: ACM, 2011, pp. 373--382.

\bibitem{LeCun1995ConvolutionalNF}
Y.~LeCun and Y.~Bengio, ``Convolutional networks for images speech and time
  series,'' 1995.

\bibitem{Likarish2009}
\BIBentryALTinterwordspacing
P.~Likarish, E.~Jung, and I.~Jo, ``{Obfuscated malicious javascript detection
  using classification techniques},'' in \emph{2009 4th International
  Conference on Malicious and Unwanted Software (MALWARE)}.\hskip 1em plus
  0.5em minus 0.4em\relax IEEE, oct 2009, pp. 47--54. [Online]. Available:
  \url{http://ieeexplore.ieee.org/document/5403020/}
\BIBentrySTDinterwordspacing

\bibitem{Liu2014}
D.~Liu, H.~Wang, and A.~Stavrou, ``Detecting malicious javascript in pdf
  through document instrumentation,'' in \emph{2014 44th Annual IEEE/IFIP
  International Conference on Dependable Systems and Networks}, June 2014, pp.
  100--111.

\bibitem{Maiorca15}
D.~Maiorca, D.~Ariu, I.~Corona, and G.~Giacinto, ``A structural and
  content-based approach for a precise and robust detection of malicious pdf
  files,'' in \emph{Proceedings of the International Conference on Information
  Systems Security and Privacy (ICISSP)}, 2015.

\bibitem{JS}
\BIBentryALTinterwordspacing
Mozilla, ``{JavaScript}.'' [Online]. Available:
  \url{https://developer.mozilla.org/en-US/docs/Web/JavaScript}
\BIBentrySTDinterwordspacing

\bibitem{PascanuMalware}
R.~Pascanu, J.~W. Stokes, H.~Sanossian, M.~Marinescu, and A.~Thomas, ``Malware
  classification with recurrent networks,'' in \emph{2015 IEEE International
  Conference on Acoustics, Speech and Signal Processing (ICASSP)}, April 2015,
  pp. 1916--1920.

\bibitem{Raff2017}
E.~{Raff}, J.~{Barker}, J.~{Sylvester}, R.~{Brandon}, B.~{Catanzaro}, and
  C.~{Nicholas}, ``{Malware Detection by Eating a Whole EXE},'' \emph{ArXiv
  e-prints}, 2017.

\bibitem{russakovsky2015imagenet}
O.~Russakovsky, J.~Deng, H.~Su, J.~Krause, S.~Satheesh, S.~Ma, Z.~Huang,
  A.~Karpathy, A.~Khosla, M.~Bernstein \emph{et~al.}, ``Imagenet large scale
  visual recognition challenge,'' \emph{International Journal of Computer
  Vision}, vol. 115, no.~3, pp. 211--252, 2015.

\bibitem{Schutt2012}
K.~Sch\"{u}tt, M.~Kloft, A.~Bikadorov, and K.~Rieck, ``Early detection of
  malicious behavior in javascript code,'' in \emph{Proceedings of the 5th ACM
  Workshop on Security and Artificial Intelligence}, ser. AISec '12.\hskip 1em
  plus 0.5em minus 0.4em\relax New York, NY, USA: ACM, 2012, pp. 15--24.

\bibitem{Shah2016}
\BIBentryALTinterwordspacing
A.~Shah, ``{Malicious JavaScript Detection using Statistical Language Model},''
  \emph{Master's Projects}, p.~70, 2016. [Online]. Available:
  \url{http://scholarworks.sjsu.edu/etd{\_}projects/476}
\BIBentrySTDinterwordspacing

\bibitem{Wael2017}
D.~Wael, A.~Shosha, and S.~G. Sayed, ``Malicious vbscript detection algorithm
  based on data-mining techniques,'' in \emph{2017 Intl Conf on Advanced
  Control Circuits Systems (ACCS) Systems 2017 Intl Conf on New Paradigms in
  Electronics Information Technology (PEIT)}, Nov 2017, pp. 112--116.

\bibitem{Wang2013}
W.-H. Wang, Y.-J. Lv, H.-B. Chen, and Z.-L. Fang, ``A static malicious
  javascript detection using svm,'' in \emph{Proceedings of the 2nd
  International Conference on Computer Science and Electronics Engineering},
  2013.

\bibitem{sdalr_wang}
\BIBentryALTinterwordspacing
Y.~Wang, W.-d. Cai, and P.-c. Wei, ``A deep learning approach for detecting
  malicious javascript code,'' \emph{Security and Communication Networks},
  vol.~9, no.~11, pp. 1520--1534. [Online]. Available:
  \url{https://onlinelibrary.wiley.com/doi/abs/10.1002/sec.1441}
\BIBentrySTDinterwordspacing

\bibitem{Wang2016}
Y.~Wang, W.~dong Cai, and P.~cheng Wei, ``A deep learning approach for
  detecting malicious javascript code,'' \emph{Proceedings of Security and
  Communication Networks}, vol.~11, no.~9, pp. 1520--1534, 2016.

\bibitem{Xu2012}
W.~Xu, F.~Zhang, and S.~Zhu, ``The power of obfuscation techniques in malicious
  javascript code: A measurement study,'' in \emph{2012 7th International
  Conference on Malicious and Unwanted Software}, Oct 2012, pp. 9--16.

\bibitem{Xu2013}
------, ``Jstill: Mostly static detection of obfuscated malicious javascript
  code,'' in \emph{Proceedings of the Third ACM Conference on Data and
  Application Security and Privacy}, ser. CODASPY '13.\hskip 1em plus 0.5em
  minus 0.4em\relax New York, NY, USA: ACM, 2013, pp. 117--128.

\bibitem{Zhao2010}
H.~Zhao and W.~Chen, ``A web page malicious script detection method inspired by
  the process of immunoglobulin secretion,'' in \emph{2010 International
  Symposium on Intelligence Information Processing and Trusted Computing}, Oct
  2010, pp. 241--245.

\end{thebibliography}

\end{document}